# Preprint Touch-less Interactive Augmented Reality Game on Vision Based Wearable Device

Zhihan Lv · Alaa Halawani ·

Shengzhong Feng · Shafiq ur Rehman ·

Haibo Li

**Abstract** There is an increasing interest in creating pervasive games based on emerging interaction technologies. In order to develop touch-less, interactive and augmented reality games on vision-based wearable device, a touch-less motion interaction technology is designed and evaluated in this work. Users interact with the augmented reality games with dynamic hands/feet gestures in front of the camera, which triggers the interaction event to interact with the virtual object in the scene. Three primitive augmented reality games with eleven dynamic gestures are developed based on the proposed touch-less interaction technology as proof. At last, a comparing evaluation is proposed to demonstrate the social acceptability and usability of the touch-less approach, running on a hybrid wearable framework or with Google Glass, as well as workload assessment, user's emotions and satisfaction.

Augmented Reality Game · Touch-less

All authors are corresponding authors.

Z. Lv
SIAT, Chinese Academy of Science in China E-mail: lvzhihan@gmail.com

A. Halawani
Tillmpad fysik och elektronik of Umeå University in Sweden
Computer Engineering and Science Dept., Palestine Polytechnic University, Hebron, Palestine
E-mail: alaa.halawani@umu.se

S. Feng
SIAT, Chinese Academy of Science in China E-mail: sz.feng@siat.ac.cn

Sh.Rehman
Tillmpad fysik och elektronik of Umeå University in Sweden E-mail: shafiq.urehman@umu.se

H. Li
Royal Institute of Technology(KTH) in Sweden E-mail: haiboli@kth.se

## 1 Introduction

Pervasive games are computer games that have one or more salient features that expand the traditional spatial boundaries of games, temporally or socially [53]. They represent an exciting development in playing games which leverages the use of sensor; visualization and networking technologies provide immerse live-action game experiences [28], and extend the game experience out into the real world [4]. The application field of pervasive game widely ranges from daily life [3] to education [50], city experience [20], cultural heritage [10], library exploration [19], mobility training [49], environmental sustainability [61] and traffic [8], for tourists [2], for office spaces [63] and game mastering [26]. Nowadays, handheld and wearable devices supported by advance in multiple sensing systems, have given rise to pervasive computing in which multimedia information is embedded into the physical world around us [7]. Meanwhile, natural interaction such as gestures, body movement, gaze and physical awareness are increasingly required [9] and they are started to be applied, yet their true potential for intuitive use is little known by people [19]. Genres of pervasive games are admittedly classified as augmented/mixed reality, pure location-based, mobile and trans-reality games [29]. Accordingly, the designed pervasive games presented in this work can be classified as both augmented reality games and mobile games, because they can be adapted to either handheld devices or head-mounted devices.

Augmented Reality is the fusion of real and virtual reality, where computer graphic objects are blended into real footage in real time. Augmented reality enhances players' immersive experience [55] [70], but the interaction with physical objects tied to the virtual control apparatus makes the complexity of game states and state functions dramatically increase [71]. Therefore, an intuitive and sensitive interaction approach is the technology demanded urgently to replace the previous technologies which draw more attention [6]. Previous researches have proved that natural human computer interaction (HCI) plays a key role in pervasive games [81] and it will also become the future trend [68].

In addition, many augmented reality games require touching and pointing inputs regardless of playing games in a mobile GUI. But the main problems they are facing are *'the occlusion problem'* and *'the fat finger problem'* [73]. Touch-less interaction can solve these problems and enhance the operation space resolution, which make it at least 5-10 times clearer than 2D touch for interaction with mobiles [78]. Additionally, another motivation frequently mentioned is that touch-less interaction identify more gestures in more natural patterns, which allows players to focus more on the games rather than on the device. Recent popular touch-less solutions provide a level of spatial input in a mobile setting [15]. A series of industrial mature products (i.e. Kinect, Leap motion) have already become popular and some novel gesture recognition approaches have been developed on these devices [75] [58], but they have not provided ubiquitous interfaces for augmented reality computing applications on smartphones or novel projected glasses yet. In addition to hands, feet and combination of hands and feet in games are used and coordinated control mechanisms of both modalities have been proven effectiveness in human-computer interaction. Feet gestures also have been considered for controlling movements on smartphones [11] and playing games [56]. It is believed that touch-less interaction approach, using hand or/and feet gestures to interact with devices, can enhance an impressive experience not only on handheld devices , but also on wearable devices.

A parallel trend, the miniaturization of mobile computing devices permits 'anywhere' access to the information [51]. Therefore, wearable device-based cameras initially present the wide applicable trend. The development on touch-less interaction methods which are well-supported by most smartphones and wearable devices, set up an easy and convenient bridge between players and games. Since the WIMP (windows, icons, menus and pointers)-based interaction techniques mentioned above are designed for desktop computers and do not support interaction on the move, as an important characteristic of wearable computers, it makes enormous sense to explore other alternative modalities, such as body gestures, to provide input to wearable devices. The touch-less interaction approach in this paper has tackled one of the most important challenges of interaction with wearable devices. Some existed games also can use the proposed system to provide touch-less interaction, such as the early pervasive games described in [23] [34].

In this paper, the potential of the touch-less approach is researched by performing 'in air' interaction gestures on two vision-based wearable devices: a hybrid wearable framework for smartphones where users mount smartphone on their wrists or knees [35], and head-wearable vision device, i.e. Google Glass. Three primitive augmented reality games have been developed which are based on the integrated real-time hands and feet detection and tracking algorithm and make user experiments with complex questionnaires. The questionnaires are designed to evaluate the usability of the eleven dynamic hand/foot gestures, workload assessment, users' emotions and satisfaction. The new contributions compared to the previous conference publications [35] [42] [43] [39] include: an overview of the need of touch-less interaction in augmented reality game; battery performance measurement; a more complete description and evaluation of the proposed algorithm; detailed gestures description; the portability research about the touch-less method running either in our designed wearable framework or with Google Glass; more complete and deeper user studies with workload and emotion tests as well as comparing evaluations between running on different devices. This paper includes seven sections: Introduction, System, Use Cases, Proof of Concept, User Study, Discussion and Conclusion.

## 2 System

The system integrates two parts based on human-centered design: (1) Vision-based wearable hardware; (2) Touch-less interaction-based augmented reality games software. When users wear the hardware, i.e. the wearable framework on their wrists or knees; the smart glasses on their heads, the software on the hardware will track the finger/foot motion (in front of the camera) to simulate the touch event with the dynamic gestures.

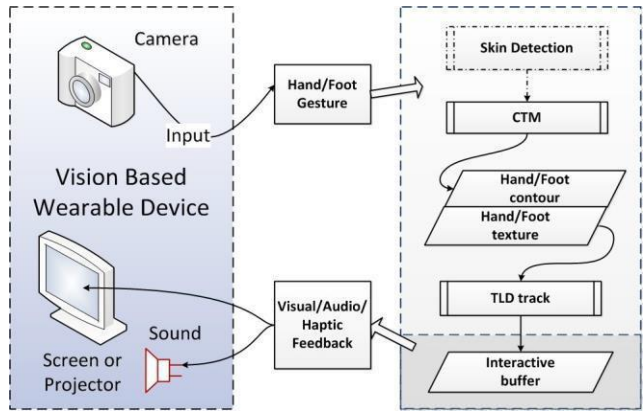

**Fig. 1** System level diagram depicting our touch-less interaction approach.

2.1 Gesture Recognition Algorithm

The touch-less interaction algorithm uses dynamic programming-based CTM algorithm [59] to localize the finger/shoe-contour for hand/foot, complemented optimally by lightweight skin detection [52], and employs TLD [27] framework to strengthen the tracking process in every frame. The touch-less interaction algorithm runs in real time, which is especially targeted at mobile devices for the realization of the interaction between human and wearable devices on the wearable framework as well as the Google Glass, enabling a wide variety of gestures.

In the first step, contour-based template matching (CTM) approach is developed to find the best region containing human finger/foot contour in the camera view. A human finger/foot template is modeled off-line and then it is compared with image frames acquired from smartphone's camera to find an optimal match as interactive feet [59] and fingerink [1] using dynamic programming in the previous work. Dynamic programming is an algorithm that is used to ensure globally optimal solutions of problems, as it always returns the highest score match. It has been proved in many state-of-the-art works [1] [17] [57] [54].

2.2 Enhanced Gesture Detecting and Tracking

The developed the contour based template matching (CTM) approach is developed by using template matching based on dynamic programming (DP) algorithm [1], in order to find the best region containing the hand/ foot contour in the camera view. A hand/ foot template is modeled off-line and then compared with image frames acquired from camera of the smartphone to find an optimal match. Following is a brief clarification of the concept. Inputs to the DP module are the binary templates and the binary edge images. Two main issues are to be considered here: template deformation and template matching.

**Algorithm 1:** The TLD algorithm with CTM localization complemented by skin detection

**Input**: Real-time video from camera
**Input**: Hand or foot contour template $T$ for $ROI_0$, where $ROI_0$ is the $ROI$ of the first frame, $L_0 = \{R_0\}$, where $R_0$ is the content surrounded by $ROI_0$ in the first frame; $L_0$ is the online model in time $t = 0$.
**Output**: current localization $ROI_t$

1  *current localization* $\leftarrow 0$;
2  **if** *isHandGesture()* **then**
3  $\quad$ *skinRegions* $\leftarrow$ *detectSkinRegions*(*videoFrame$_t$*)
4  *gestureContour* $\leftarrow$ *cbTemplateMatching*(*skinRegions, templates T*)
5  -Dynamic programming based match using $T=\{(p^{11},p^{12},\cdots,p^{1M}),(p^{21},p^{22},\cdots,p^{2M}),\ldots,(p^{N1},p^{N2},\cdots,p^{NM})\}$, where p represent the density of every point of the template
6  -$ROI_0 \leftarrow$ Bounding box BB=$\{$ top-left($x^{min},y^{min}$), bottom-right($x^{max},y^{max}$)$\}$ of the best match, where $x^{min}, y^{min}, x^{max}, y^{max}$ are the minimum and maximum values of the x and y in given frame
7  *gestureTexture* $\leftarrow$ *getGesture*(*gestureContour*)
8  **while** *active input feed from camera in time t* **do**
9  $\quad$ Tracking
10 $\quad$ Detection $R$ in online model $L_{t1}$
11 $\quad$ Learning
12 $\quad$ -$P - N$ Experts
13 $\quad$ -$Lt \leftarrow L_{t1} \cup$ *Positive samples from growing.*
14 $\quad$ -$Lt \leftarrow L_{t1}$ ¥ *Negative samples from pruning.*
15 $\quad$ Integrator
16 $\quad$ -Validation
17 $\quad$ -$R_t \leftarrow$ *the most confident evaluated result ROI$_t$ $\leftarrow$ bounding box of $R_t$*

### 2.2.1 Skin Detection

The performance of the skin detection algorithm depends on many factors such as light conditioning, skin color etc. The skin detection employs HSV color space of skin images to obtain the skin regions. An acceptable range of light and skin color is provided and it works well in the experimental cases. Only the extreme lighting has problem. The detected skin regions are used to instantiate hypotheses of fingers in order to enable real-time hand segmentation even on resource-restraint devices.

### 2.2.2 Contour-based Template Matching

- Deformable Templates
  DP is an algorithm that is used to ensure a globally optimal solution of a problem, as it always returns the highest score match. It has proven to be successful in many contributions [18]. Hence, we decided to adapt the DP algorithm to the task of shoe/foot detection. Namely, we use DP to search for the best fit of a simple shoe template in a given image.
  If treated as a rigid entity, the template will be unusable since it would not tolerate different shoe/foot shapes and transformations. Instead, it is divided into shorter segments that are allowed to move within a certain range during the matching process. Deformation is introduced as a result of the controlled segment movements.

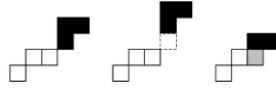

**Fig. 2** Examples of deformation caused by segment movements. Enlargement using 1-pixel gap (*middle*). Shrinkage using 1-pixel overlap (*right*).

The example in Figure 2 illustrates the concept. It is assumed that the template is divided into several 3-pixel segments. The left column of the figure shows the original spatial arrangements of two segments in the template. By allowing the segments to move one pixel, a set of deformations can be achieved by introducing a gap (middle column) or an overlap (right column) between the segments. Segments can be shifted to any position in the image to be searched, provided that the relative displacement between two consecutive segments is not greater than one pixel. I.e., if two consecutive segments are shifted by $o = (o_x, o_y)$ and $p = (p_x, p_y)$ respectively, then relative displacement is governed by:

$$|o - p| = \max \left( |o_x - p_x|, |o_y - p_y| \right) \leq 1. \tag{1}$$

The degree of template flexibility is then governed by the segment length. If each segment contains 3 pixels, then an overall shrinkage or enlargement of around $1/3 = 33\%$ can be introduced. A mixture of gaps and overlaps will result in a set of possible deformations.

– Template Matching The Viterbi DP algorithm is used for the matching process. If viewed as a trellis, each column corresponds to a segment, and nodes in that column correspond to possible shift values of that segment in the image. Arcs connecting nodes in two consecutive columns are governed by Equation 1. We search for the path through the trellis that maximizes the accumulated score, $R$. For segment $i$ shifted by $p$, $R$ is given by:

$$R(p, i) = \max_{p' \in \sigma(p)} \left\{ R(p', i-1) \right\} + V(p, i), \tag{2}$$

$$\sigma(p) = \{p' \in \Omega : |p' - p| \leq 1\}.$$

$\Omega$ is the set of all possible shifts. $V(p, i)$ is the local reward given to node $(p, i)$ and it is equal the number of edge pixels segment $i$ will cover if placed at position $p$. When $R$ has been calculated for the last segment, the algorithm backtracks the optimal path starting from the node with the highest $R$ value in the last column.

### 2.2.3 CTM-TLD

CTM approach is an indispensable step for interactive hand (two fingers outstretched) detection in the beginning because TLD cannot be initialized by a generic hand/foot(shoe) model. Moreover, the classification learning and training methods cannot detect hands/feet, which may not have distinctive features. Hand template is modeled off-line in advance and then it is compared with image frames acquired from skin regions to find an optimal match. CTM allows a slight deformation of template, which allows it to successfully detect the approximate gestures of hands and feet of different people. CTM provides TLD a region of interest (ROI) for initialization (As in Figure 1).

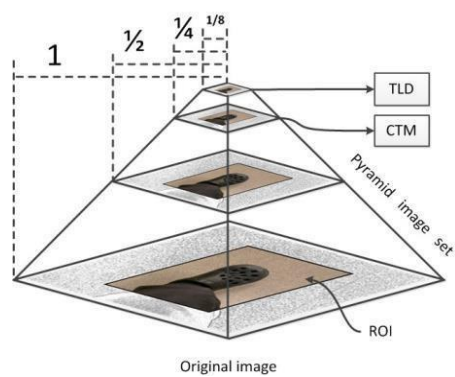

**Fig. 3** Pyramid structure for accurate and real time results.

In order to select the most suitable tracking algorithm in the second step, three tracking algorithms are ported to the Android smart phone and the performance and efficiency are compared. TLD [27], HoughTrack [21], PixelTrack [13] are three current advanced tracking algorithms. In the case, HoughTrack is the most robust and time-consuming algorithm, while PixelTrack is the fastest and also algorithm with good performance. However, both algorithms build look-up tables in the initial step, which costs more than 10 seconds on the 1.5GHz smartphone. Therefore, TLD algorithm is adopted, which costs less than 1 second in the initial step but still gets a high score in the tracking accuracy test on the smart phone. In addition, TLD can support scale varying that is one of the characteristics of spatial motion interaction. TLD algorithm combines elements of tracking, learning and detection in the 2D image space to make it a long-term tracker. The TLD tracker uses a tracking strategy of overlapping blocks and tracks every block by Lucas-Kanade optical flow method. TLD reduces the image processing time of the pure CTM. Besides, a limitation of occasional false detection exists in CTM if the hand/foot gestures exceed the restriction of deformation or complex backgrounds interferes detection rate. The proposed combined touch-less algorithm successfully solves all the unexpected detection results of hand/foot gestures. The enhanced algorithm is followed as in Algorithm 1. In this paper, the algorithm is combined into one core library with adaptation of the wearable framework as well as the Google Glass and its parameters are adjusted to enable detection and tracking for both hand and foot flexibly. A series of further user studies is done, including social acceptability, usability, work load, user emotions and user satisfaction.

### 2.2.4 Optimization

To improve the efficiency of the proposed algorithm, the size of the first frame is scaled down to 25% for foot and 12.5% for hand and afterward CTM is applied.

From the second frame the image is scaled to 12.5% for foot and 6.25% for hand, and then TLD tracking is employed (As in Figure 3). The scale ratio is different for hand or foot because the distance between the cameras of the wearable devices with hand is closer than the foot (As in Figure 9). Both the 'hand target test' and 'foot target test' results indicate that both efficiency and success rate of the tracking are close to 100%.

2.3 Technical Evaluation

*2.3.1 Computational Performance*

The algorithm execution is real time (i.e. >10fps) and accurate even when the hand gesture is moving fast and in image blur cases. It is worth mentioning that the method can detect the designed dynamic hand-gestures as long as the correct hand texture is detected in the DP step. To verify the accuracy of the algorithm, it has been tested on the hand-gesture-video data set containing 120 mobile video sequences (each has more than 1000 frames) of single hand-gesture and foot-gesture for various sizes and orientations in the acceptable range (i.e. 33% deformation)(see Figure 4). Compared to the previous limited test data set [42], the new collected data set used in this paper contains video-sequences from clustered office as well as outdoor lighting conditions. The new data set has considered more real environments that the system could meet during running. The proposed algorithm successfully locates and tracks the hand-gesture in almost all videos. The success rate overall gesture recognition is 99.76% which is noted for private video sequences. The *'finger target test'* (i.e. a visual inspection to check how well the algorithm has localized the users' fingertip in every video frame) results indicates that both efficiency and success rate of the tracking are close to 100%.

The algorithm execution is real time (i.e. >10fps) and accurate even when the hand/foot gesture is moving fast and in image blur cases. It's worth mentioning that the method can detect all kinds of hand/foot gestures used in our games as long as the correct hand/foot texture is detected in the CTM step (As in Figure 5).

Our touch-less approach also solved some common limitations of vision-based hand gesture interaction for PC. For example, one of the limitations is that the hand is hard to be detected when the user is trying to point to the lower part of the screen. When the fingers are pointed to the lower part of the screen, the hand is hardly visible and hard to be recognized by the hand detection algorithm. To solve this problem, touch-less interaction mode in this algorithm is adopted to avoid the direct hand contact on the screen and increase the resolution of interaction. Meanwhile, the ROI selection method in the pyramid-based optimization process can expand or shrink the recognition region freely so that the hand gesture close to the border of the camera can be recognized.

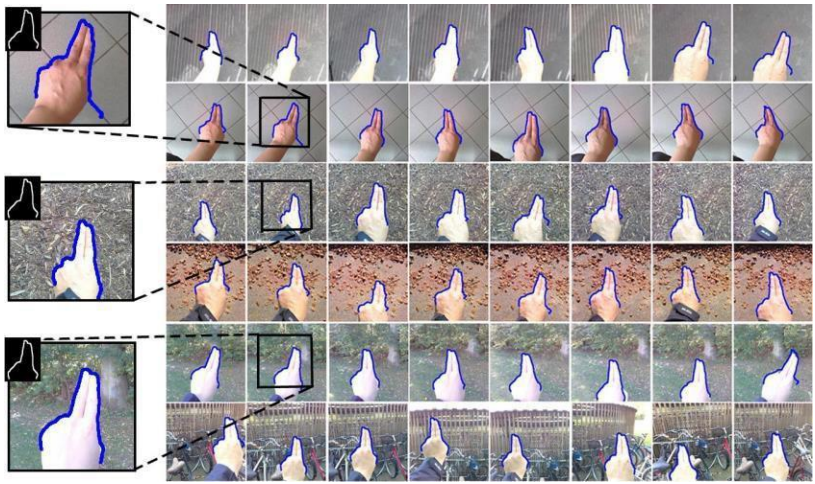

**Fig. 4** Modeled template and accurate detection of user's finger gesture. Results for template based gesture detection algorithm applied on our dataset.

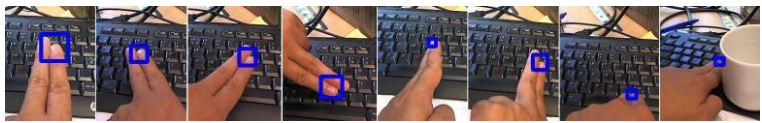

**Fig. 5** The tracking results for a variety of finger gestures with background interference

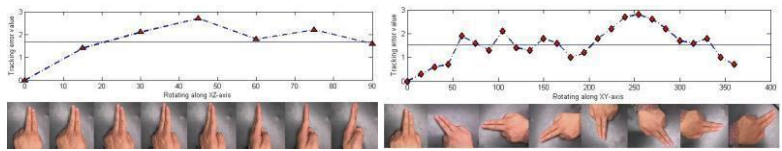

**Fig. 6** The 'sample user' gesture tracking error in pixels as Left when rotating fingers along XZ-axis and as Right when rotating fingers along XY-axis.

### 2.3.2 TrackingAccuracy

To test the accuracy in tracking performance of the algorithm, a 'precision test' of finger motion in XY and XZ axis are performed. For this test, the calculated error is in pixel per 15-degree motion. The sample user's finger tracking error in XY axis (with circular motion) is shown in right Figure 6. These error values are recorded while the finger is rotated along each 15 degrees. The test is repeated for twenty users and the mean value is computed. The mean value of the error during the finger motion is 1.516 pixels with sd = 0.8572. The maximum error value is 2.8 pixels at 255 degrees and finally, the error value is 0.7 pixels at 360 degrees. Similarly, the tracking error when the tracking object is rotated along XZ axis is shown in left Figure 6. The mean error during the motion is 1.6857 pixel with sd = 0.7255. The maximum error value of 3 pixels is recorded when the 'finger motion speed' is altered.

*2.3.3 Battery Performance*

In order to evaluate the impact on battery performance of the touch-less interaction technology, a series of trials are conducted by using identically configured HTC one X smartphones. The measurement is not conducted on Google glasses since Google glasses sometimes shut down automatically due to overheating, which can interrupt the measurement. Using the designed touch-less gestures, the power consuming of two scenarios are compared: 1) touch-screen interaction, and 2) touch-less interaction. During the comparison of the battery life, the interferences like wireless antenna are turned off. Each smartphone runs uninterrupted for ten minutes, which is not so for most context-aware applications [12], but useful in order to provide multiple group case analysis in short time.

Our findings indicate that touch-less gestures have a measurable impact on battery life. Simply using the touch-less interaction drains the battery by 20%. Yet for almost every gesture, using the framework brings improved efficiency over using a single device. 'Swing finger to left' and 'swing finger to right' gestures draw especially more attention in the trials because they are commonly used by handinteraction applications. However, to achieve meaningful power saving with gestures, the cost of computing needs to be higher than the real-time images capturing. To demonstrate this, the camera is kept open but no any hand or foot in the camera view, and it is found that the system saves 12% power than computing gestures when hands or feet appear in the camera view. Overall, the battery trials have provided us with an insight. Interaction's energy efficiency improves as no gestures in the camera view.

## 3 Use Cases

This section introduces the designed hands and feet motion sensing gestures. After that, it will compare the interaction technology running on three devices through intuitive spatial analysis. The core element of this section is to show the different situations that the proposed system can be used (e.g. when a player is standing, walking or sitting etc). The proposed system can be applied on some existing augmented reality games and to be used to provide touch-less interaction, such as those described in [23] [34].

### 3.1 Motion Sensing Gesture

Pervasive games can bring the pleasure of the games to ordinary life [53], which means the players can play the pervasive games with their usual postures in the life, such as standing, sitting or lying [43]. Due to its high performance and efficiency, the interface is able to sense hands/feet-based motion interaction. A linear method is used to calculate the foot motion as shown in figure 7, where $D$ is the distance of the movement, $T$ is the time consuming of the movement and $\alpha$ is the tilt angle of the movement on behalf of the movement direction which can be calculated according to relation between the movement on X and Y axis. Velocity $V$ can be calculated fast by $V = D/T$.

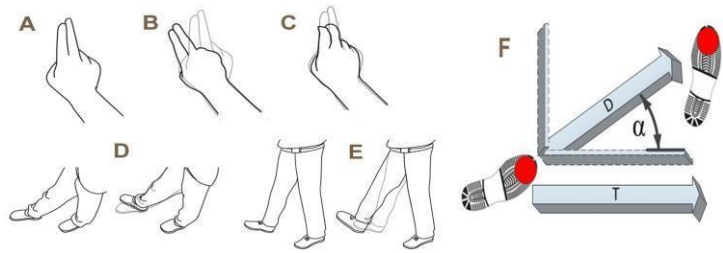

**Fig. 7** From up-to-down: Hand gesture; Foot gesture; Foot Motion Sensing Computing chart.

Figure 7 shows the basic gestures of hands $(A, B and C)$ and feet $(D, E and F)$. $A$ shows the initial hand gesture and orientation and B illustrates that the finger swings from left to right (it can also swing the opposite direction). Except defining the direction by swinging fingers, users can also move the visual mouse by moving their hands. $C$ represents the flex and extension motion of fingers, which is similar to the gesture of clicking when people use mouse. $D$ indicates the motion of feet, which can trigger some events related to the position in certain scenes by tiptoe. $E$ illustrates the forward motion which can trigger optional events kicking motion and moving motion according to the moving speed of users' feet. The estimation of speed is computing with the linear algorithm shown in the sub-figure $F$. Next section will discuss the motion sensing gesture interaction method of running on different devices.

## 3.2 Running on Handheld Devices

Figure 8 shows that, regardless of the body gesture (sitting as in $A$, $B$; standing as in $C$, $D$ or lying as in $E$), the device can recognize the hands or feet gesture via the rear camera of the device, according to which the device can manipulate the software. Besides, the user can also handle it with the front camera $(F, G)$. Red zone indicates the users' sight, yellow zone indicates the camera's video capture zone. Apparently, the distance from eye to screen has no difference, but the distance from the camera to hand $(A, C, F and G)$ is clearly shorter than that from camera to feet $(B, D, E)$. At the same time, it can be noticed that the gesture of hands or feet interacting with the device via the rear camera is displayed through the screen, presenting to the users indirectly. While for those gestures via the front camera, users can see the visual reactions in the screen directly. In addition, one hand has to hold the handheld device $(A, B, C, D, E)$, which will weaken the interaction efficiency, although the hand holding the device can still make some incomplete touch interaction. $F$ and $G$ show that in those scenes, the hands are totally free. However, the mobile devices in those scenes require to be supported by another object which also makes it inconvenient to control by feet.

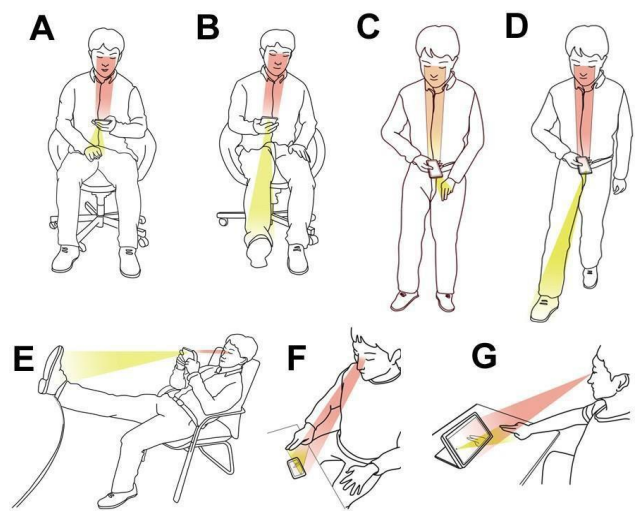

**Fig. 8** The players can play the touch-less interaction pervasive games on either smartphone or tablet in a variety of postures; i.e., sitting, lying, and/or standing.

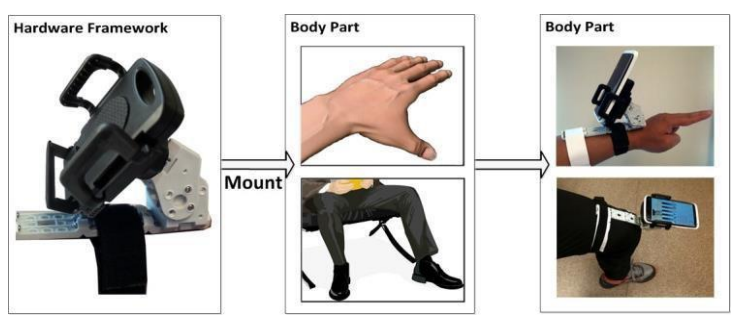

**Fig. 9** The wearable framework mounts on wrist for hands (eg., finger) interaction and on knee for feet interaction.

### 3.3 Running on Hybrid Wearable Framework

The hardware of the hybrid wearable framework is comprised of three main hardware components: mobile phone holder, holder base and fastener strip. The connection between mobile phone holder and holder base is an active loose leaf, which can be twisted to adjust the pitch angle of the smartphone camera for adaptation of wrist and knee. The framework can be fixed on wrists or knees firmly by the fastener strip for detecting of hand and foot gesture interaction respectively, as shown in Figure 9.

### 3.4 Running on Smart Glasses

Google glass has many impressive characteristics (i.e. voice action, head wake up, wink detection) and those are defined as human-glass interface (HGI) technologies.

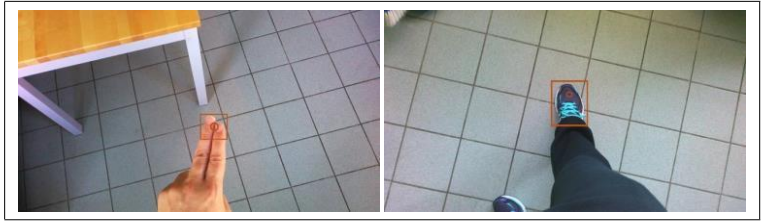

**Fig. 10** Hand and foot gesture recognition on Google Glass.

Google glass is free of the occlusion problem and fat finger problem, which frequently occur in direct touch-controlled mode for once and always. However, Google Glass only provides a touch pad that includes tactile sensing with simple gestures such as *'tapping and sliding your finger'* and this tactile sensing is a unidimensional interaction in fact. Instead of the traditional two-dimensional interaction based on a complete touch screen of smartphone, the unidimensional *'swipe the touchpad'* interaction with a row of *'Cards'* replacing conventional two-dimensional icon desktop, has limited the intuition and flexibility of HGI. Therefore, implementing a vision system of light-weight gesture recognition is a new challenge, in which cameras capture real-time videos of the users' movements and then certain algorithm is applied to calculate and determine what the gestures mean, no necessity for users to hold any device. Captured by ADB command, the screenshot of the implementation on the Google Glass is shown in Figure 10. The demonstration has been proved [41] [40].

How the user wearing the Google glass interacts with the glasses through the glass camera capable to recognize the gestures of hands and feet is illustrated in Figure 11. User's sight indicated by the red zone basically coincides with the camera captured zone indicated by the yellow zone. This is because the eyes are close to the camera of glasses and they have the same orientation. Thus, the gestures seen with user's naked eyes are on the whole the same as those captured by the camera, which makes this kind of interaction via wearable glasses more intuitive than that by virtue of mobile camera. Besides, this kind of interaction frees up the users' hands completely so that operating the device with both hands and their combination is well-supported (shown as *C*). In this scene, since hands and feet are in the same side, manipulating the device with hands and feet combination is also supported. *D* and *E* indicate that, benefiting from the ubiquitous feature, this kind of light-weight interaction technique will surely find wide application on many smart devices such as smart glasses or smartphones in the near future. Users are able to connect with each other via the Wi-Fi or bluetooth network, and influence each other in the same virtual world with their own limbs, for example, playing a football game in the virtual field as in *E*.

## 4 Proof of Concept

Three primitive augmented reality games based on the wearable hardware system are presented in this paper. These games are designed to prove that the touch-less interaction system can play a role in augmented reality games on a variety of use cases introduced in the section of use case. These games demonstrate the usefulness, viability and flexibility of the touch-less interaction approach which is applied to investigate its possibility and limitation, to illustrate and explore different use cases. 'Bouncing ball game' and 'Football game' have the similar game rules. 'Foot-Play Piano' proves the capability of touch-less interaction approach to control rhythm by foot. Accordingly, a user study comparing social acceptability and usability, workload assessment, uses' emotions and satisfaction is reported.

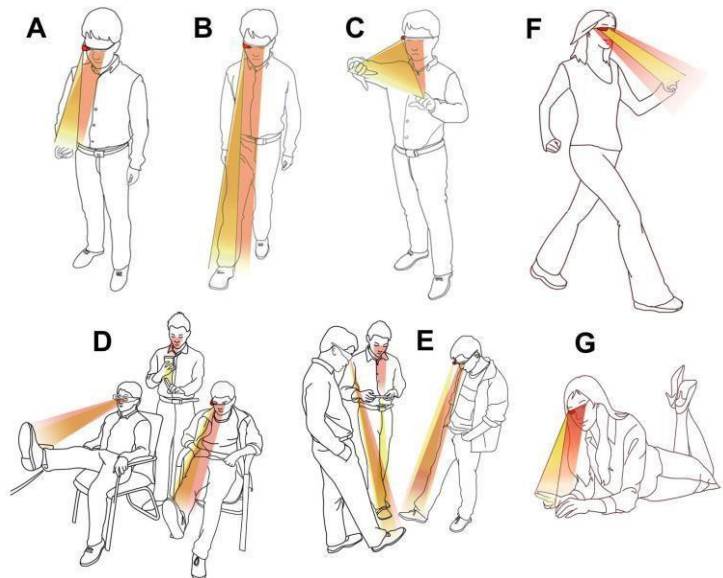

**Fig. 11** The touch-less interaction on smart glasses.

The augmented rendered glove/shoe's image or color-marked cursor overlaid on the camera view on the screen provides the user with direct and immediate reference of the actual hand position relative to the screen. This user is expected to perceive the touch-less approach as being stretched from his or her body (i.e. hand, foot) when manipulating on the screen. In order to complete this interaction loop, our system maps the physical hand/foot onto the screen coordinates. When the user manipulates the icon/button on the screen by touch-less gestures, the system calculates and updates the position of the augmented rendered image/cursor overlapping with the screen mapped from the location of the gesture accordingly. Touch-less algorithm is able to recognize a number of user gestures including simple actions such as *'Swing Finger'* and complex operations such as *'Finger Flexion and Extension'*. Since the names of all the designed gestures are to some extent self evident, we have not made detailed explanation of each gesture in this paper. All augmented reality games are implemented in C++ and Java with Android SDK/NDK and the images processing tasks are realized by using OpenCV.

A bouncing ball game based on the touch-less interaction approach is developed on Android platform as a proof of the support of speed and orientation-based movement gesture of the concept of 'touch-less'. The scene of the game is a handball court with a camera view overlay. An augmented reality glove image is rendered following the finger gesture. The handball always bounces back and forth on the court area until it's into the goal or the player uses the *'Collide the ball'* gesture to intercept and hold it. When the player holds the ball, he/she can use the *'Swing Finger Slowly'* gesture to dribble the ball to the desired location. *'Swing Finger Fast'* gesture can be used for throwing or bunting the ball, at the same time, a red arrow at the top-right corner of the screen is the index of the speed value by degree of elongation. In order to compare the effects of UI and pure graphic UI of the augmented reality rendered interaction, a *'smile face'* button in the bottom right of the screen is designed to trigger the hide/show status of the camera view, shown in the left of Figure 12.

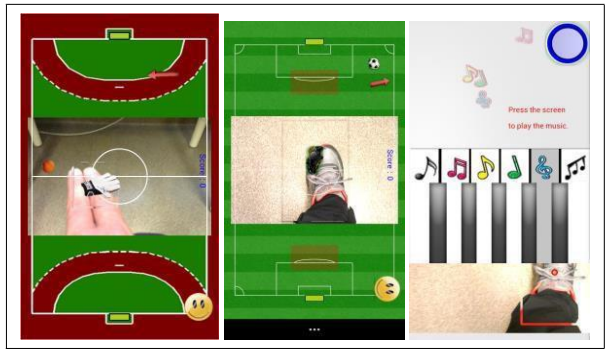

**Fig. 12** From left to right: Bouncing ball Game. Football Game. Foot-Play Piano App

As a real-time multi-modal augmented reality game, football game presents the game graphics and status information by virtue of smart devices video, audio and vibration and by employing an augmented reality image rendered technology to create a vividly visual football game to the users. In the system, the players use real foot to kick the ball by using *'Kick The Ball'* gesture. When the player holds the ball, he/she can use the *'Move Foot Slowly'* gesture to dribble the ball to the desired location. *'Move Foot Fast'* gesture can be used to throw or bunt the ball, shown in the middle of Figure 12 middle. The Bouncing ball game and the Football game have the same game rules, respectively for hand and foot interaction.

'Foot-Play Piano' represents the application scenario of user study. The goal is to demonstrate the rhythm control of touch-less interaction through the beat of music. An opt ion is given in this game to play music on the users' android smartphones by using foot gestures. There is a circle progress plate on the top right corner of the screen which is used to show how long the user presses the piano key, eg., a circle is one Mora; semi-circle is a half Mora. When the start button is pressed, the collection of foot movements through the android camera is started automatically. When a collision with any augmented piano-key by *'Press The Key'* gesture is detected, the code is related to that key, shown in the right of Figure 12.

## 5 User Study

The record of neutral views expressed by the evaluators may indicate technical flaws or even suggest essential script, usability and technical improvements [30]. Tools related to pervasive games that allow players to record their subjective experiences during an ongoing game have been developed [69]. In the process of evaluation, a choice has been made by us to utilize the traditional statistic method. Two user groups participated in the experiment are used in the comparative research. In order to evaluate the usability of the gestures respectively, we set a user group consisted of 15 volunteer participants (6 female included) from a diverse background in the age from 25 to 46 (m=30.13, sd=5.74). They are provided with the touch-less interaction devices and tested on interaction with the apps on the hybrid wearable framework, and then they were required to answer some open-ended questions related to the new interaction mode. All the participates have personal smartphones with touch screen but none of them has any experience with touch-less interfaces on smartphones before. First of all, the moderator introduces the concept of 'wearable' via demonstration. A two-stage questionnaire is designed by us (and the design of the experiment) to obtain participants' reactions to this modality: first, Social Acceptability of the Augmented Reality Games; second, Usability of the Gestures; third, User Workload Assessment; fourth, User Emotions; fifth, User Satisfaction. In order to confirm that the proposed touch-less on the smart glasses can bring about better user experience, a new user group including 15 participants are invited to have a test on Google Glass, and then we compares the scores of the two groups. Only the designed gestures in the user study are compared instead of the experience of whole games. The second user group consisted of students, researchers, and teachers with different professions and their ages are from 24 to 41 (m=28.733, sd=4.079). Both of the two user groups are volunteers interested in performing user tests based on the novel interaction technology. The user study is described in this section and corresponding results will be discussed in the next section.

### 5.1 Social Acceptability

Regarding significant research, we have tried to understand engagement in games by measuring player experience [5]. After measurement, the first stage is comprised of the same questions as that inspired by Rico et al. [60], planning to elicit social acceptability of both augmented reality games. Rico et al. distinguishes social acceptability by two factors: audience (alone, partner, colleague, friend, family, stranger) and location (home, passenger, workspace, pub, street, driving).

A two-sub-stage questionnaire is designed by us so as to evaluate the social acceptability of the augmented reality games. First of all, the participants watch a video showing the tutorial performing each gesture of both games. Then, the first sub-stage questionnaire is carried out and participants are invited to try the device in person. In the second sub-stage, a smartphone is installed on the wrist and the knee of each participant and three augmented reality games are performed. Afterward, the other

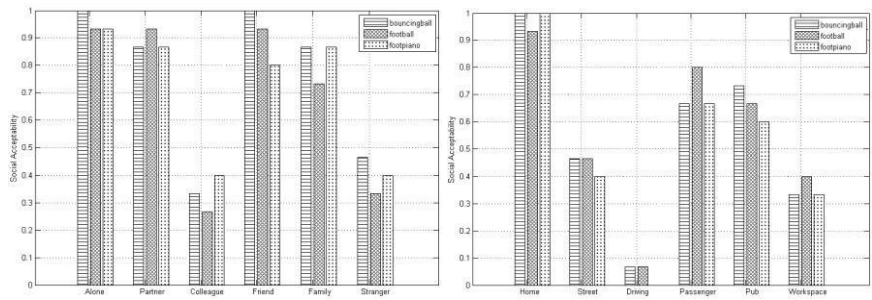

**Fig. 13** Social acceptability rating by locations and augmented reality games.

sub-stage questionnaire is implemented. Measurements come out in the form of both questionnaires.

The rating of mean acceptance among all contexts is above 0.5 for each pervasive game ($m_{bouncingball} = 0.6500$; $m_{football} = 0.6222$; $m_{footpiano} = 0.6056$). These results are illustrated in Figure 13 left and Figure 13 right. The Kruskal-Wallis (KW) test ($\chi^2 = 9.19, p = 0.0269$) reveals a significant effect on social acceptability for the augmented reality games at $p = .05$ level, followed by pair-wise comparisons reveal that *'football'* is a bit more acceptable than *'footpiano'*. ($meanrank_{bouncingball} = 358.5, meanrank_{football} = 348.5, meanrank_{footpiano} = 342.5$).

5.2 Usability

Usability differs from user satisfaction and user experience because usability also considers usefulness. The second stage is comprised of Likert scale questionnaire to reveal the usability of the eleven designed gestures. These gestures include six hand touch-less gestures (*'Swing Finger to Left'*, *'Swing Finger to Right'*, *'Finger Flexion and Extension'*, *'Swing Finger Slowly'*, *'Swing Finger Fast'*, *'Collide the ball'*) in the one game based on hand interaction, and five designed gestures (*'Kick The Ball'*, *'Move Foot Slowly'*, *'Move Foot Fast'*, *'Press The Key'*, *'Release The Key'*) in the two games based on foot touch-less interaction. It is no surprise that these interaction metaphors found their way into modern body-interaction-enabled games on devices ranging from smartphones and tablets to large tabletop displays. We believe these metaphors will be well established for the wearable framework based on touch-less interaction.

In the test for the first user group, the mean of usability score reveals that all gestures are applicable. The standard deviation of the usability of the *'Swing Finger Slowly'* gesture is much larger than those of others, indicating that there is fewer consensus among participants on the usability of this gesture in Table 1.

The KW test reveals a significant effect ($\chi^2 = 24.63, p = 0.0061$) on usability for the gestures at $p = .01$ level. The mean-rank reveals that gesture *'Finger Flexion and Extension'* gets the highest score and *'Swing Finger Slowly'* gets the lowest score on the aspect of usability.

**Table 1** Mean and Standard Deviation of usability of gestures for the two groups

|  | $m_1$ | $sd_1$ | $m_2$ | $sd_2$ |
|---|---|---|---|---|
| Swing Finger to Left | 4.2000 | 0.7746 | 4.3333 | 0.7237 |
| Swing Finger to Right | 4.0667 | 0.7988 | 4.3333 | 0.7237 |
| Finger Flexion and Extension | 4.6000 | 0.5071 | 4.6000 | 0.6325 |
| Swing Finger Slowly | 3.4000 | 1.1212 | 3.6667 | 0.8997 |
| Swing Finger Fast | 4.1333 | 0.9155 | 3.7333 | 0.8837 |
| Collide the ball | 4.2000 | 0.7746 | 4.3333 | 0.7237 |
| Kick The Ball | 3.4667 | 0.6399 | 3.6000 | 1.0556 |
| Move Foot Slowly | 3.5333 | 0.9904 | 3.8000 | 1.1464 |
| Move Foot Fast | 3.6667 | 1.0465 | 3.8667 | 0.9904 |
| Press The Key | 4.0667 | 0.7988 | 4.2000 | 0.9411 |
| Release The Key | 3.7333 | 1.0328 | 3.9333 | 1.2228 |

**Table 2** Wilcoxon signed rank test result for gestures on wearable framework or google glass

| Wilcoxon signed rank test | p-value | H | signedrank |
|---|---|---|---|
| Swing Finger to Left | 0.5000 | 0 | 3 |
| Swing Finger to Right | 0.2500 | 0 | 6 |
| Finger Flexion and Extension | 1 | 0 | 3 |
| Swing Finger Slowly | 0.3594 | 0 | 21 |
| Swing Finger Fast | 0.2500 | 0 | 2 |
| Collide the ball | 0.7500 | 0 | 10 |
| Kick The Ball | 0.7813 | 0 | 13 |
| Move Foot Slowly | 0.2500 | 0 | 6 |
| Move Foot Fast | 0.2500 | 0 | 6 |
| Press The Key | 0.6250 | 0 | 7.5 |
| Release The Key | 0.6250 | 0 | 7 |

The means of usability scores in the test for the second user group are higher than that for the first user group except 'Swing Finger Fast' gesture. The KW test reveals a significant effect ($\chi^2 = 18.64, p = 0.0451$) on usability for the gestures at $p = .05$ level. It reveals that the designed gestures are not restricted by the device exhaustively.

Wilcoxon signed-rank test [74] in Table 2 is used to compare two groups' scores, since subjective data is ordinal, not interval, and hence parametric tests should not be used. Here we notice our sample size n = 15. The value h = 1 indicates that the test rejects the null hypothesis that there is no difference between the grade medians at the 0.05 significance level. While the value h = 0 indicates that the test fails to reject the null hypothesis that there is no difference between the grade medians at the 0.05 significance level.

### 5.3 Workload Assessment

The third stage is NASA-TLX workload assessment. The NASA-TLX workload assessment items are rated low (as in Figure 14). However, it is noticeable that the answers for the foot have higher values on all parts. In particular, 'frustration' with the foot for 'Foot piano' game has almost twice value as that for the 'Bouncing ball' game using hand. Similar to the answers from the questionnaire, the majority of the users find that the mental demand and physical demand of the games are quite low, and feet is a bit more frequently used than hands.

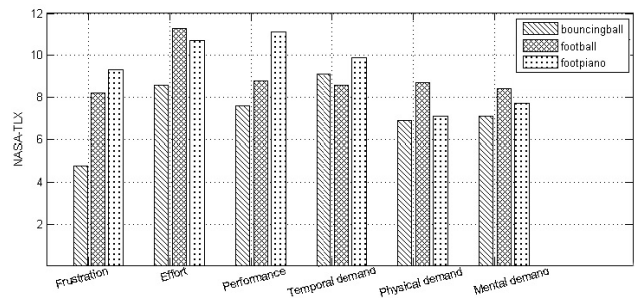

**Fig. 14** NASA-TLX workload results for the three augmented reality games. The values are from 1 (Low) to 20 (High).

**Table 3** Wilcoxon signed rank test result for emotions on wearable framework or google glass

| Wilcoxon signed rank test | p-value | H | signedrank |
|---|---|---|---|
| Hand Pleasant | 2.4414e-04 | 1 | 0 |
| Hand Unpleasant | 6.1035e-05 | 1 | 120 |
| Hand Low Control | 0.3662 | 0 | 67.5 |
| Hand High Control | 0.9265 | 0 | 58 |
| Foot Pleasant | 0.0016 | 1 | 5.5 |
| Foot Unpleasant | 6.1035e-05 | 1 | 120 |
| Foot Low Control | 0.9873 | 0 | 45 |
| Foot High Control | 0.9082 | 0 | 43.5 |

### 5.4 Emotions

Emotion is a significant part of users' decision-making ability. Emotional response is part and parcel for usability. Emotion test is able to evaluate the product against different emotions that a user goes through. Happy emotion denotes good user experience and sad emotion denotes bad user experience. By this way, we can directly communicate with the stakeholders on how the product i s generated on emotion test and better decisions facilitated.

The analysis on the players' emotional responses can provide pleasant game experiences to successfully design augmented reality mobile games [14]. In the fourth stage, in order to assess the users' emotions such as arousal and happiness based on the hand and foot interaction, the Geneva Emotions Wheel (GEW) is applied [62]. GEW allows us to deal with the pleasantness and control the dimension of emotions and is handed to the users immediately after they conducts the given tasks hand and foot. Hence it can be ensured that the emotions that have been experienced during the interaction with devices are still persistent. The GEW is constructed in a way that it represents four different combinations of characteristics about control and pleasant-ness. The Figure 15 depicts such collected values where the pleasant attributes are rated with 2.6, unpleasant attributes with 0.7 for the hand, and 2.5 and 1.2 for the foot on the wearable framework. At the same time, the pleasant attributes are rated with 3, unpleasant attributes with 0.2 for the hand, and 3 and 0.3 for the foot on Google Glass. In other words, the result reveals obviously an more positive emotion because the pleasant score is much higher than the unpleasant one. Meanwhile, Google Glass brings more positive emotions to the users than the wearable framework. The second dimension of the GEW refers to the feeling of control over the emotion. The average values of High Control situations and Low Control ones are 2.0 and 1.2 for the hands, and 2.0 and 1.1 respectively for the feet on the wearable framework. Meanwhile, the average values of High Control and Low Control are 2.0 and 1.1 for the hands, and 2.0 and 1.2 correspondingly for the feet on the wearable framework.

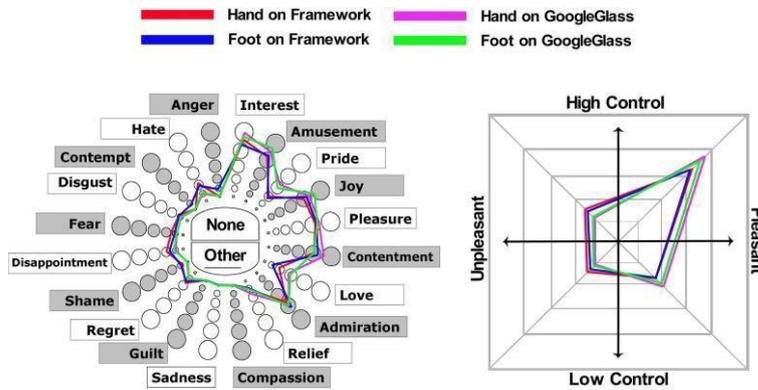

**Fig. 15** User's average emotional ratings. The Geneva Emotion Wheel shows ratings (Left) organized by emotion and (Right) averaged across quadrants.

The result of Wilcoxon signed-rank test [74] in Table 3 reveals that smart glass is of influence on the user's pleasant emotions for using the designed gestures.

### 5.5 Satisfaction

**Table 4** Mean and Standard Deviation of the user satisfaction parameters for 'Bouncing Ball' (i.e. $m_{bb}$, $sd_{bb}$), 'Football Game' (i.e. $m_{fg}$, $sd_{fg}$) and 'Foot Piano' (i.e. $m_{fp}$, $sd_{fp}$) Games.

|  | GUI | Comfort | Accept | Interest | Willingness |
|---|---|---|---|---|---|
| $\mathbf{m}_{bb}$ | 6.0000 | 5.0000 | 6.0909 | 6.3636 | 4.8182 |
| $\mathbf{sd}_{bb}$ | 0.7746 | 0.5071 | 1.1212 | 0.9155 | 0.7746 |
| $\mathbf{m}_{fg}$ | 5.0000 | 4.4545 | 5.2727 | 5.7273 | 5.0000 |
| $\mathbf{sd}_{fg}$ | 1.6733 | 1.6348 | 1.3484 | 1.1909 | 1.8439 |
| $\mathbf{m}_{fp}$ | 4.5455 | 4.0000 | 4.0909 | 4.7273 | 4.8182 |
| $\mathbf{sd}_{fp}$ | 1.2933 | 1.0000 | 1.3003 | 1.1037 | 1.2505 |

In the fifth stage, User Satisfaction refers to the comfort and acceptability of the system for its users and other people affected by its use [16]. The '*Effectiveness*' is about whether a task could be accomplished with the specified system or not. As all users have completed the task within the given time, the effectiveness of the system is proven. The '*Efficiency*' is about how much effort will be required in order to accomplish the task. An efficient system ought to require as little effort as possible. The 'algorithm execution time' is considered as an indicator of efficiency in the experiments which are real-time. The reaction time of a user can be used as an indicator of efficiency as well. Both 'Effectiveness' and 'Efficiency' will be considered in details in future studies. 'User satisfaction' is the only factor being concerned regarding the experiments in this work. A useful measurement mode of user satisfaction can be made if the evaluation team's measurement is based on observing users' attitudes towards the system. Thus, it is possible to measure users' attitudes through conducting a questionnaire. The questions used for measuring user satisfaction are given as follows:

**GUI** - Is the Graphical User Interface is easy to use?
**Comfort** - Is this application comfortable to use?
**Accept** - Is this application acceptable?
**Interest** - Is this application interesting?
**Willingness** - Are you willing to pay to buy such a product?

For this subjective questionnaire, the Likert style 7-point rating system is used, which scale i from 1 to 7. The test results are illustrated in Table 4.

## 6 Discussion

The result of test on social acceptability shows that participants are inclined to play *'bouncingball'*, *'football'*, *'footpiano'* in the open (pub). Moreover, the social acceptability reveals the augmented reality games enjoys seldom popularity in the work environment(colleague, workplace).

The result of usability test indicates that Google Glass basically brings about more intuitive and comfortable user experience. The decreased score of 'Swing Finger Fast' gesture can be explained as that the head-wearable device isn't fixed on the wrist or knee so that more joints can be used in hand motion, thus 'Swing Finger Fast' gesture requires more strenuous effort. Another presentation is that 'Swing Finger to Left' gets more scores than 'Swing Finger to Right' in the first test, but gets the same scores in the second test. The reason should be similar as previous one: in the first test, the wearable framework is fixed on the left wrist so that the users feel more comfortable to swing the hand to the left than right. However, the available joints are not restricted in the wrist in the second test. As a result, the users shows no care regarding using left or right hand any more. The result of Wilcoxon signed-rank test reveals that smart glass impose no influence on the designed gestures, meaning that the gestures are applicable to ubiquitous context, in other words, the gestures are suitable for augmented reality games. The result of workload test shows that users are inclined to dispatch priority to the hand so as to achieve the same mission due to the less mental and physical demands.

The situation of emotion test indicates, in analogy with the result of pleasantness, that people tend to have control over the emotion while the difference is that Google Glass hasn't brought significant improvement on controlling these proposed touch-less gestures. In general, the proposed interaction method with both devices affects users' emotion significantly. Positive emotion as that depicted on the right side of the GEW are regarded slightly more effective than negative emotion as t h a t depicted on the left side.

With regard to the measurement on users' satisfaction, participants get very good scores on both game, indicating that the touch-less interfaces are the valid substitutes for present touch-based mobile devices. Experimental result also shows that the participants consider touch-less interfaces interesting and comfortable to be used. Among other parameters, Comfort and Willingness get fewer scores. Further investigation on these less scores indicates that the users are in need of time to get used to 'Arm-bound' interfaces and they are not sure about the price of this new technology. Overall rating of foot touch-less interaction is lower than that of hand touch-less interaction, which shows that using wearable devices to handle the hand interaction challenge bears more acceptability and usefulness than foot interaction.

Concerning the use cases, presented three augmented reality games have been completed in order to conduct user study deep by comparing these unconventional interaction styles with established ones and to identify the most promising augmented reality games among the wide range of novel interaction possibilities. User studies confirm that the touch-less interaction technology is intuitive and controllable and that mappings between gesture and manipulation on mobile phones. Since there is no necessity for the approach of touch-less motion interaction we proposed to depend on range sensor or dual camera which is more expensive and less convenient, it is suitable for modern devices using smart glass. The touch-less motion interaction on head-wearable devices using smart glass extends the resolution of operation space not only because it widens the distance between hands/feet and camera, but also that it completely frees the users' hands from the handheld devices. By this way, both hands are available to create more gestures. In addition, the angle of view and the position of camera with frontal glass are close to the right eye, and the visual feedback from the glass projector is directed at the right eye. And that is intuitive WYSIWYG interaction and benefit can be obtained from the coordination between head and hands/feet gestures.

# 7 Conclusion

A series of designed gestures and the evaluation research related to touch-less interaction technology on vision-based wearable device are proposed in this paper. The proposed touch-less interaction technology offers a full view of body control mode on wearable devices and that is the difference from traditional interaction technology. By comparing result of user studies, it can be revealed that the smart glass is a preferable platform for touch-less interaction approach. Although they are fun to use, present touch-less games based on augmented reality on wearable devices have certain limitation. For example, their performance are limited to hands and feet gestures. Normal full-body gestures are even not considered in the research version.

This paper provides a low-cost solution for the professional augmented reality games on mobile phones, as well as popular wearable devices such as projected glasses, or even Memoto and GoPro. The hybrid wearable framework demonstrated in this paper is possibly applicable to be worn on other body parts. Moreover, it is possible to replace it with other types of sensors to improve the touch-less interaction modality of the wearable framework. Capacitive, electronic-field, ultrasound, and IR-based sensors may be included since these sensors may be capable of generating a hand/foot image from the raw data. The proposed touch-less interaction technology has been planed to apply in many fields range from everyday entertainment and communication [36] [79] to traditional research fields such as geography [48] [65], ocean [64] [46], city [45] [33] [32] and biology [66] [47], clinical assist [37] [38] [44]. The new network data management algorithms [25] [72] [24] and advanced image processing algorithms [77] [80] [76] will be considered to improve the performance. In addition, 'GroupMe' system [22] and 'FRAP' framework [67] are considered to be employed as the sensing system of group-aware smartphones to support group management and activity organization in augmented reality games. We believe that it's entirely possible to perform multi-modal and multi-user tasks and communication through using proposed touch-less interaction approach. Finally, a plan is made to conduct research on the emotion evaluation of game players in long-term [31].

**Acknowledgments.**

The authors are thankful to Muhammad Sikandar Lal Khan for the preliminary hardware device support, to Liangbing Feng for kind help at SIAT and to our friends for their fruitful discussions and code-sharing.